\documentclass[sigconf]{acmart}
\AtBeginDocument{%
  }

\copyrightyear{2025}
\acmYear{2025}
\setcopyright{cc}
\setcctype{by}
\acmConference[CHI '25]{CHI Conference on Human Factors in Computing
Systems}{April 26-May 1, 2025}{Yokohama, Japan}
\acmBooktitle{CHI Conference on Human Factors in Computing Systems (CHI
'25), April 26-May 1, 2025, Yokohama,
Japan}\acmDOI{10.1145/3706598.3713131}
\acmISBN{979-8-4007-1394-1/25/04}

\usepackage{xcolor}
\usepackage{soul}

\begin{document}

\title{Does the Story Matter? Applying Narrative Theory to an Educational Misinformation Escape Room Game}

\author{Nisha Devasia}
\email{ndevasia@uw.edu}
\affiliation{%
  \institution{University of Washington}
  \city{Seattle}
  \state{Washington}
  \country{USA}
}

\author{Runhua Zhao}
\email{runhz@uw.edu}
\affiliation{%
  \institution{University of Washington}
  \city{Seattle}
  \state{Washington}
  \country{USA}
}

\author{Jin Ha Lee}
\email{jinhalee@uw.edu}
\affiliation{%
  \institution{University of Washington}
  \city{Seattle}
  \state{Washington}
  \country{USA}
}

\renewcommand{\shortauthors}{Devasia et al.}

\begin{abstract}
    Rapid spread of harmful misinformation has led to a dire need for effective media literacy interventions, to which educational games have been suggested as a possible solution. Researchers and educators have created several games that increase media literacy and resilience to misinformation. However, the existing body of misinformation education games rarely focus upon the socio-emotional influences that factor into misinformation belief. Misinformation correction and serious games have both explored narrative as a method to engage with people on an emotional basis. To this end, we investigated how 123 young adults (mean age = 22.98) experienced narrative transportation and identification in two narrative-centered misinformation escape room games developed for library settings. We found that propensity for certain misinformation contexts, such as engagement with fan culture and likelihood to share on social media platforms, significantly affected how participants experienced specific measures of narrative immersion within the games. We discuss design implications for tailoring educational interventions to specific misinformation contexts. 
\end{abstract}



\maketitle

\section{Introduction}
The ubiquity of the internet and social media platforms have made the task of creating and disseminating information easier than ever before. Subsequently, the ease of information distribution has led to a rise in the prevalence of misinformation. Misinformation (false or misleading information) and disinformation (false information spread with malicious intent) have many widespread and often harmful effects on society due to their ability to shape people's beliefs and behaviors \cite{Ecker_Lewandowsky_Cook_Schmid_Fazio_Brashier_Kendeou_Vraga_Amazeen_2022}. These range from increases in vaccine-preventable diseases \cite{Garett_Young_2021, Lee_Sun_Jang_Connelly_2022} to political polarization \cite{Ribeiro_Calais_Almeida_Meira-Jr_2017}. This has led to calls to feature misinformation more predominantly in mainstream media literacy curricula \cite{Dame-Adjin-Tettey_2022}. Media literacy was shown to positively correlate with correct determination of the accuracy of online information \cite{Kahne_Bowyer_2017}. However, current curricula do not adequately address all of the skills and strategies necessary to navigate the rapidly evolving information landscape \cite{Kellner_Share_2019}. 

Games have been suggested as a promising educational medium for effective media literacy interventions \cite{Chang_Literat_Price_Eisman_Gardner_Chapman_Truss_2020, Contreras-Espinosa_Eguia-Gomez_2023}. The immersive nature of games gives players a safe space to investigate complex issues \cite{Schulzke_2014} such as the spread of misinformation. In particular, serious games, which often integrate learning outcomes with engaging mechanics \cite{Molnar_Kostkova_2013}, offer a unique and interesting platform for imparting critical thinking skills to students growing up in the information age. Indeed, researchers and educators have created games that aim to improve media literacy \cite{Chang_Literat_Price_Eisman_Gardner_Chapman_Truss_2020, Contreras-Espinosa_Eguia-Gomez_2023, Kiili_Siuko_Ninaus_2023}, and effectively inoculate players from misinformation and disinformation \cite{Basol_Roozenbeek_Linden_2020, Maertens_Roozenbeek_Basol_van-der-Linden_2021, Roozenbeek_Van-Der-Linden_2019, van-der-Linden_Maibach_Cook_Leiserowitz_Lewandowsky_2017}. However, there are limitations to the existing body of game-based misinformation interventions. Many are strictly focused on educational outcomes, presented as quizzes with elements of gamification \cite{Contreras-Espinosa_Eguia-Gomez_2023}. While informative, these interventions primarily address the rational processes of misinformation correction. In reality, the processing and subsequent adoption of misinformation is also heavily influenced by psychological drivers and personal belief \cite{Ecker_Lewandowsky_Cook_Schmid_Fazio_Brashier_Kendeou_Vraga_Amazeen_2022}. It is equally essential for designers of misinformation education games to facilitate player exploration of the socio-emotional influences that can lead to the acceptance and spread of misinformation, an area which has been underexplored in the media literacy space \cite{wedlakegames}.

The fields of misinformation correction and serious games have both explored a common method to engage with people on an emotional basis: narrative. Reading, processing, and identifying with narratives is a fundamental component of how we organize our interpretations of reality \cite{Bruner_1990}. Misinformation research has found that narratives which evoke strong emotions in readers can have a corrective effect on misinformed opinions \cite{Cohen_Tal-Or_Mazor-Tregerman_2015, Ophir_Romer_Jamieson_Jamieson_2020, Sangalang_Ophir_Cappella_2019}. Game studies have also shown that narratives can be powerful tools for empathizing strongly with characters and mediating beliefs \cite{Domínguez_Cardona-Rivera_Vance_Roberts_2016, Iten_Steinemann_Opwis_2018,Mahood_Hanus_2017}, which can further lead to learning outcomes within the scope of educational games \cite{Abdul_Jabbar_Felicia_2015, Jackson_O’Mara_Moss_Jackson_2018, McQuiggan_Rowe_Lester_2008, McQuiggan_Rowe_Lee_Lester_2008, Naul_Liu_2020}. However, despite the effectiveness of narrative persuasion in both misinformation correction and educational games, current misinformation games are notably lacking in narrative-driven learning mechanisms, as their primary focus tends to be on improving skill-based or knowledge-based information literacy. 

To this end, we sought to investigate different measures of narrative persuasion within the context of two narrative-based misinformation education games developed at University of Washington. The games themselves are identical in educational content, and only differ in terms of their narrative framing. The first game, \textit{The Euphorigen Investigation} (hereafter referred to as Euphorigen), was designed for a general audience by our core research team during the COVID-19 pandemic \cite{Cho_Coward_Lackner_Windleharth_Lee_2023}. It was piloted in libraries across the country, and following its enthusiastic reception, librarians expressed a desire for more games with different narrative themes and topics intended for the different communities they work with. This inspired the creation of the second game, \textit{The Galaxy} (hereafter referred to as Galaxy), which was co-designed with fans of a large Kpop (Korean pop) group to create a game that appeals to online fan communities. The narrative measures used to compare these two games in this study are \textit{transportation} (level of immersion in the world of the story) and \textit{identification} (level of emotional and cognitive connection with a character in the story), two key components of narrative's ability to change people's behavior \cite{Green_Brock_2000, Moyer-Gusé_2008, Cohen_2017}. We posed the following research questions:
\begin{itemize}
    \item \textbf{RQ1}: How do the target audiences of each game experience narrative transportation and identification as compared to the non-target audiences?
    \item \textbf{RQ2}: What factors affect narrative transportation and identification in these misinformation education games?
    \item \textbf{RQ3}: How did participants connect the game narratives to misinformation in their own lives?
\end{itemize}

We recruited 123 participants from the University of Washington, and split them into two groups to play the two games. In addition, participants were further split by how closely they aligned with fan culture relevant to Galaxy, resulting in four different study groups. After participants finished playing the game, we measured their levels of narrative transportation \cite{Green_Brock_2000} and identification \cite{Cohen_2017} with the game, and gained a qualitative understanding of how they connected the game to their own lives through a post-game debrief. Our quantitative analysis showed that certain patterns of social media usage, such as engagement with fan culture and propensity for sharing on social media, were predictive factors for transportation and identification within the games. Qualitatively, we explored how participants interact with misinformation in their daily lives, and investigated potential reasons for vulnerability to different types of misinformation. We discuss the implications of tailoring educational interventions to specific misinformation contexts, and advocate for a greater focus on the narrative framing of such interventions. 

\section{Related Work}
\subsection{Narrative Theory}
Narratives are capable of `transporting' individuals, a psychological process in which they become immersed in the world of the narrative. Transportation has been described as a key mechanism underlying the effect of narratives on individuals’ attitudes and beliefs \cite{Green_2008}. Another related phenomenon is that of identification, “a process that consists of increasing loss of self-awareness and its temporary replacement with heightened emotional and cognitive connections with a character” \cite{Cohen_2017}. Busselle \& Bilandzic utilize transportation and identification to describe an individual's deictic shift into a story, in which they "shift the center of their experience from the actual world into the fictional world and position themselves within the mental models of the story", enabling them to assume the points of view presented by the narrative \cite{Busselle_Bilandzic_2008}. Moyer-Gusé's model posits that transportation and identification reduce reactance to topics in the narrative that the individual would find controversial outside of the storyworld, reduce counterarguing with ideas that they may normally disagree with, and reduce selective avoidance to arguments that contradict one’s existing beliefs and attitudes \cite{Moyer-Gusé_2008}. These behaviors keep the individual consistent with the attitudes and behaviors adopted during the deictic shift into the narrative. Slater \& Rouner's extended Elaboration Likelihood Model \cite{Slater_Rouner_2002}, which builds from Petty \& Cacioppo's Elaboration Likelihood Model \cite{petty2012communication}, suggests that the cognitive processing of narratives suppresses resistance to persuasive messages contained within the story. The effectiveness of persuasive messaging within a narrative was found to be associated with the degree of transportation into the story and identification with the characters \cite{Green_Brock_2000, Slater_1997}, which led Slater \& Rouner to further argue that transportation and counterarguing are mutually exclusive. If a message recipient is able to counterargue the information transmitted by the narrative, this implies that they were not sufficiently transported. In previous work, Slater \& Rouner found that narrative messages were more persuasive than factual arguments, particularly for participants with pre-existing attitudes that countered the persuasive messaging in question \cite{Slater_Rouner_1996}. In accordance with the extended Elaboration Likelihood Model (e-ELM), there is evidence that narratives can completely overwrite preexisting attitudes regarding controversial issues \cite{Igartua_Barrios_2012, Slater_Rouner_Long_2006}, which may be of particular relevance to misinformation contexts. This study draws from the theoretical grounding provided by the e-ELM and uses validated constructs developed by Green \& Brock  \cite{Green_Brock_2000} and Cohen \cite{Cohen_2017} to investigate how participants experienced narrative transportation and identification within the misinformation game context.

\subsection{Misinformation Education and Games}
Media literacy games that specifically focus on misinformation are relatively few in number, and narrative-centered misinformation games are less common still \cite{Devasia_Lee_2024}. In a recent review, we found that only 11 of 37 identified digital misinformation education games qualified as narrative driven \cite{Devasia_Lee_2024}. Contreras-Espinosa \& Eguia-Gomez identified 24 games with explicit learning outcomes for media literacy, with their creation ranging from 2012 to 2023 \cite{Contreras-Espinosa_Eguia-Gomez_2023}. They draw from the report Get Your Facts Straight!: Toolkit for Educators and Training Providers \cite{Get-Facts_ALL-DIGITAL-Week_2020}, which includes a goal-oriented methodology and curriculum that breaks down media literacy education on misinformation and fake news into three main learning areas \cite{Contreras-Espinosa_Eguia-Gomez_2023}. Kiili et al. performed a systematic overview of games aimed at misinformation education specifically, which lists a set of 15 games designed primarily by researchers studying critical media literacy. Many of these games cite \textit{inoculation theory} as their theoretical basis \cite{Kiili_Siuko_Ninaus_2023}. Inoculation theory is a particularly widespread technique in misinformation media literacy, especially in misinformation games. It exposes people to the tactics and flawed argumentation used to spread misinformation in order to `immunize' them against similar attempts faced in the future \cite{Cook_Lewandowsky_Ecker_2017, van-der-Linden_Maibach_Cook_Leiserowitz_Lewandowsky_2017}. The success of the award-winning \textit{Bad News} game \cite{Roozenbeek_Van-Der-Linden_2019} and its predecessors \textit{Go Viral!} \cite{basol2021towards} and \textit{Breaking Harmony Square} \cite{roozenbeek2020breaking}, developed by researchers at the Cambridge Social Decision-Making Lab, likely popularized inoculation theory as a design approach. However, recent findings suggest that games based on inoculation theory may simply increase the likelihood of conservative reporting - that is, when presented with several comparable true and false news items, participants who played \textit{Bad News} or \textit{Go Viral!} were more likely to report the item as fake overall \cite{modirrousta2023gamified}. Recent research has looked beyond inoculation based approaches into more social misinformation game environments; in particular, escape rooms \cite{Cho_Coward_Lackner_Windleharth_Lee_2023, paraschivoiu2021escape, buchner2024playing, pun2017hacking}. Educational escape rooms, or EERs, integrate multiple perspectives into problem-solving, which makes them a particularly valuable medium through which to explore misinformation \cite{tercanli2021educational}. We utilize EERs both in this study and in our misinformation education research more broadly \cite{Cho_Coward_Lackner_Windleharth_Lee_2023, wedlakegames}. 

In their review of educational lenses regarding the current information ecosphere, Barzilai \& Chinn (2020) list four framings for educators and researchers to consider \cite{barzilai2020review}, and offer broader perspectives for efforts to support media literacy. These are: 1. not knowing how to know, which education might mitigate through inoculation approaches; 2. fallible ways of knowing, by teaching how to cope with cognitive biases and limitations; 3. not caring enough about the truth, by cultivating dispositional intellectual virtues, and 4. disagreeing about how to know,  by acknowledging and coordinating multiple epistemologies. As inoculation approaches (lens 1) may increase conservative reporting rather than critical engagement with misinformation \cite{modirrousta2023gamified}, the games used in this study take a different approach, primarily focusing on using narrative to aid reflection on information behaviors, as well as exploring socio-emotional and sociocultural literacy through educational escape rooms (lenses 2 and 3).

\subsection{The Psychology of Misinformation} \label{psych}
The drivers of misinformation belief can be divided into two categories: cognitive and socio-emotional \cite{Ecker_Lewandowsky_Cook_Schmid_Fazio_Brashier_Kendeou_Vraga_Amazeen_2022}. Understanding these distinct drivers is crucial for developing effective strategies to counter misinformation. Research suggests that rational attitudes are best addressed with rational messages, whereas emotional beliefs are best changed with emotional messages \cite{nelson2002role, See_Petty_Fabrigar_2008}. Pre-existing beliefs are a strong predictor for adoption of misinformation that aligns with said beliefs \cite{Ecker_Lewandowsky_Cook_Schmid_Fazio_Brashier_Kendeou_Vraga_Amazeen_2022}; several studies have shown that people actively seek out information confirming their beliefs (\textit{confirmation bias}) and ignore dissenting information \cite{Zhou_Shen_2022, Zollo_2019}. Indeed, counterarguing against an attempted misinformation correction can strengthen an individual's belief in it \cite{Ecker_2017}. To this end, counter-narratives have been used as debunking measures to deconstruct strongly held beliefs \cite{White_2022}, such as those common among smokers \cite{Ophir_Romer_Jamieson_Jamieson_2020, Sangalang_Ophir_Cappella_2019}. Evoking a strong emotional response and identification with the main character of such counter-narratives was shown to have mediating effects on misinformed beliefs \cite{Cohen_Tal-Or_Mazor-Tregerman_2015, de2012identification}. Notably, narrative correction was more effective for participants who strongly identified with the character \cite{Ophir_Romer_Jamieson_Jamieson_2020}. We draw on these approaches in the design of this study by measuring narrative identification with misinformation narratives within a game context. Indeed, despite the importance of both cognitive and socio-emotional drivers of misinformation, literacy approaches in current misinformation games primarily focus on the former, e.g., inoculation theory. There is limited exploration of the latter \cite{wedlakegames, Devasia_Lee_2024}, which we investigate further in this study in tandem with how narrative might influence engagement with these factors.

\subsection{Games and Empathy Building}
Game studies have shown that perspective taking in virtual environments increases empathy \cite{Estrada-Villalba_Jacques-García_2021}. The interactivity provided by games leads to personal agency, allowing players to make meaningful choices within the scope of the game world \cite{Consalvo_Busch_Jong_2019, Iten_Steinemann_Opwis_2018, Yin_Xiao_2022}. The agency to make in-game choices leads to immersion and the identification that players feel with game characters \cite{Yin_Xiao_2022}, which further affects their in-game actions and self-perception \cite{Happ_Melzer_Steffgen_2013, Hefner_Klimmt_Vorderer_2007}. Players are capable of feeling deep emotional attachment to  and identification with characters in narrative games \cite{Bopp_Müller_Aeschbach_Opwis_Mekler_2019, Hefner_Klimmt_Vorderer_2007, Sierra-Rativa_Postma_Van-Zaanen_2020}. This increases situational empathy for that character, regardless of their morality \cite{Happ_Melzer_Steffgen_2013, Iten_Steinemann_Opwis_2018}. Players exhibit a preference for actions consistent with their in-game role, termed the mimesis effect \cite{Domínguez_Cardona-Rivera_Vance_Roberts_2016}. Players who play as a morally good hero are more likely to exhibit pro-social behaviors in the game world, while those playing as the villain are prone to antisocial behaviors \cite{Happ_Melzer_Steffgen_2013}. These behaviors are in alignment with Moyer-Gusé's model \cite{Moyer-Gusé_2008}, which would suggest that players roleplay in accordance with their character to reduce cognitive dissonance. Role-playing games, which are distinguished by their strong narrative focus, are capable of inducing transportation in players \cite{Moyer-Gusé_Mahood_Brookes_2011}. The degree of transportation that a player experiences affects how strongly they feel about the morality of their in-game actions \cite{Mahood_Hanus_2017}. Game environments also provide a medium for players to understand other players with whom they may not necessarily identify with closely in real life \cite{Burgess_Jones_2021}. For this reason, games are a promising medium for misinformation interventions, and empathizing with others was one of the key design goals for the games used in this study \cite{Cho_Coward_Lackner_Windleharth_Lee_2023}. In addition, the game is intended to promote empathy for people who fall for and share misinformation in an attempt to combat the third person effect, a phenomenon in which people tend to perceive that mass media messages have a greater effect on others than on themselves \cite{corbu2020they}. Another key design principle of our games is that narrative-centered educational games can also spur attitude change. In Jackson et al.'s \cite{Jackson_O’Mara_Moss_Jackson_2018} review of narrative-centered educational games, they found that skill acquisition (measured in 33 out of 130 reviewed studies) and attitude change (measured in 15 out of 130 reviewed studies) were the most effective educational outcomes within the scope of these games. This presents an opportunity for designers of misinformation education games to not only allow for skill-building, but to also engage in the attitude changes required to address false beliefs.

\section{Methods}
To understand how players experience transportation and identification in misinformation games with different narratives, we ran 19 in-person sessions with two versions of a misinformation escape room game designed at our university. Our goal was to investigate possible differences in these measures between players who reported low vs. high engagement with certain misinformation contexts, and how those differences might inform future design of similar educational experiences. 

\subsection{Participants and Recruitment} 
\begin{figure}
  \centering
  \includegraphics[width=\linewidth]{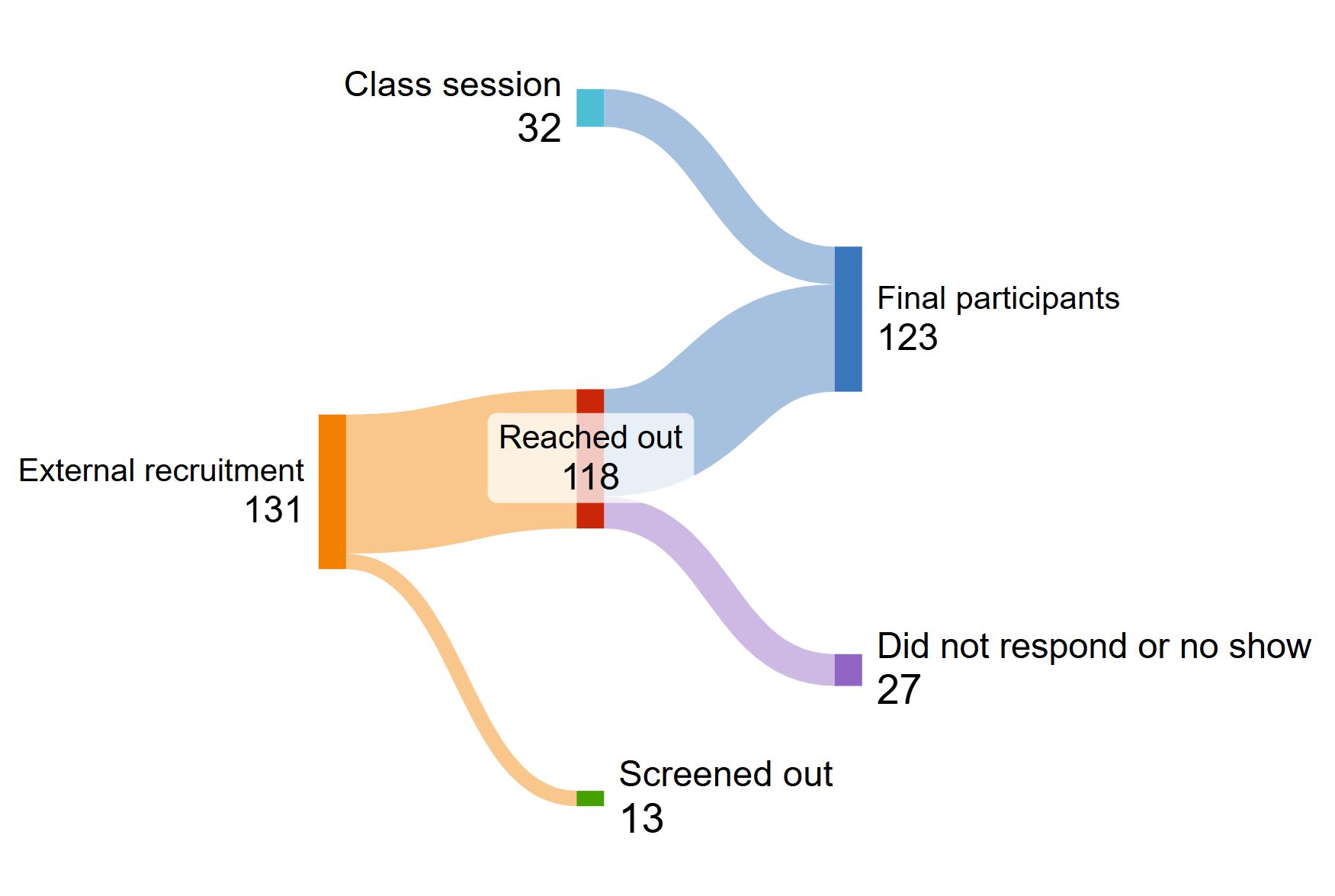}
  \Description{A Sankey diagram showing the flow of the recruitment process, including key stages and transitions.}
  \caption{A Sankey diagram representing the recruitment process. }
  \label{sankey}
\end{figure}

123 participants ranging from 18-47 years old were recruited for this study (84 female, 32 male, 7 non-binary). The average age of the participants was 22.98 $\pm$ 4.474. Multiple races and ethnicities were represented: White (26\%), Asian (60.2\%), Multiracial (5.7\%), Black or African American (4.9\%), Middle Eastern or North African (2.4\%), and Hispanic or Latinx (0.8\%). We recruited 91 of our 123 participants via a screening survey (see Section \ref{screening}) disseminated through fliers, word of mouth, and posts to gaming servers at the university. There were a total of 131 responses collected through these means, of which 13 were determined to have incomplete screening information. We reached out to the remaining 118 participants inviting them to participate, but 27 either did not respond or were no-shows to the game session, leaving 91 participants who attended one of 18 sessions. The 19th and final session was run in a class of 32 students (who were given the same screening survey to determine their study condition), completing the total of 123 participants (see Figure \ref{sankey}).

At the beginning of the session, all participants were asked to fill out a consent form detailing the data that would be collected during the session. The recruitment materials, study protocol, and data collection protocol were reviewed and approved by the University of Washington's Institute Review Board. All participants were compensated with a \$40 gift card.  

\subsection{Screening Survey} \label{screening}
The screening survey collected the following information about participants:
\begin{itemize}
    \item Demographics, namely age, gender, and ethnicity.
    \item Frequency of social and news media usage, measured on a Likert scale from 1-6 where 1 was `Never' and 6 was `Several times a day'. 
    \item Political affiliation, measured on a Likert scale from 1-7 where 1 was `Very liberal' and 7 was `Very conservative'. 
    \item Fan culture engagement, measured on a Likert scale from 1-5 where 1 was `Not at all' and 5 was `Very often'. Participants were also asked to provide popular figures/groups they engaged with, if any. 
\end{itemize}

Responses to the questions about engagement with fan culture were used to split players into the study conditions (see Section \ref{conditions}). 

\subsection{Game Descriptions}
The University of Washington's Center for an Informed Public and GAMER (Game Research) Group have created several educational experiences focused on building resilience to misinformation. In this study, we used two escape room games, Euphorigen and Galaxy. Both games share the same set of puzzles and mechanics, and their only difference is the narrative scenario. 

\begin{figure}
  \centering
  \includegraphics[width=\linewidth]{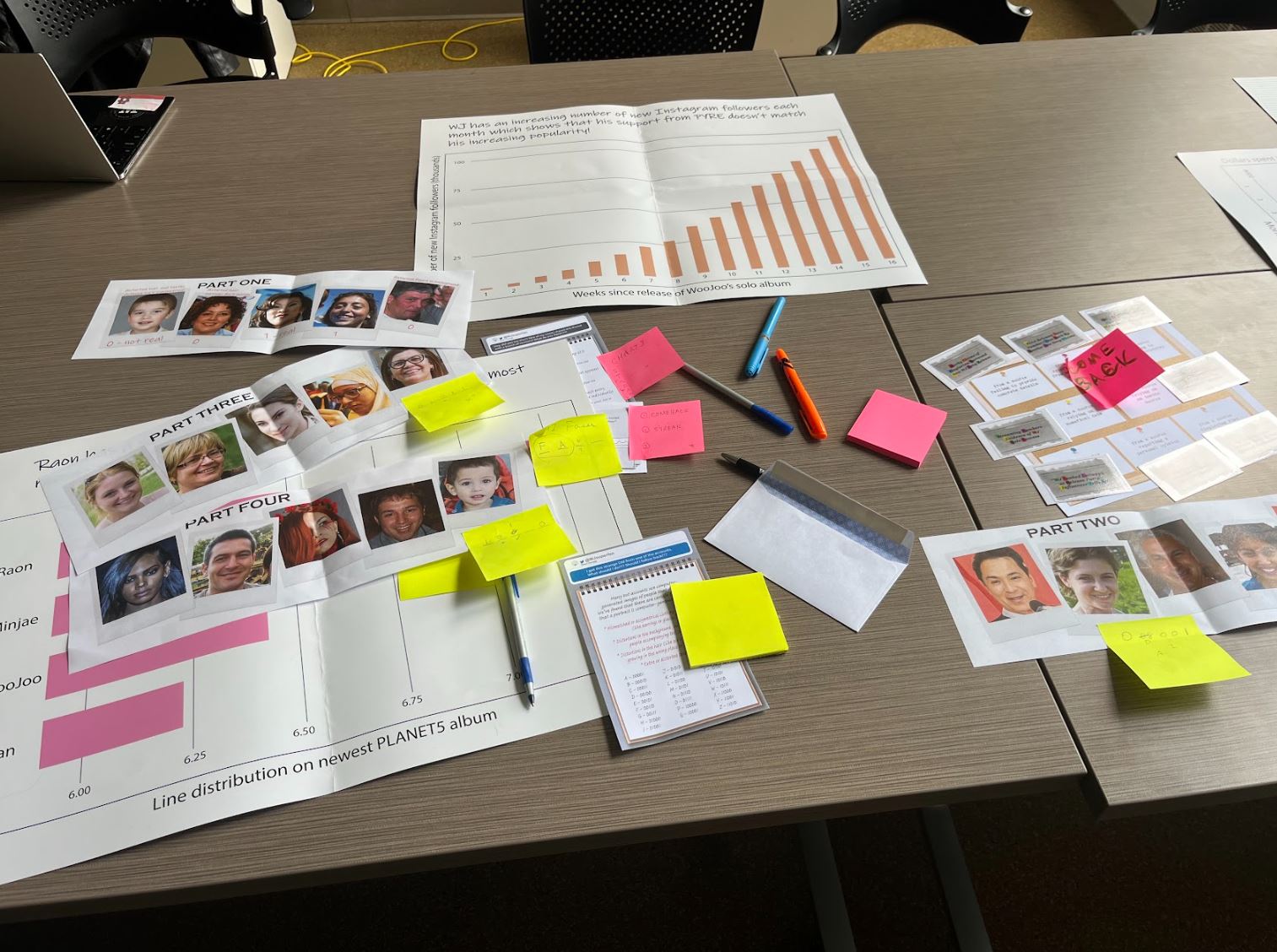}
  \Description{A representative example of the player workspace, showcasing the first three puzzles of the game.}
  \caption{A representative example of the player workspace, showcasing the first three puzzles of the game. The large bar charts (top middle and bottom left) ask players to identify misleading chart titles; the headline puzzle (right middle) asks player to identify the description matching the nature of each of the headlines; the AI face puzzle (left middle) asks players to determine which faces are AI-generated. Sticky notes and markers were provided for note-taking.}
  \label{gameplay}
\end{figure}

\subsubsection{Euphorigen} 
\begin{figure}
  \centering
  \includegraphics[width=\linewidth]{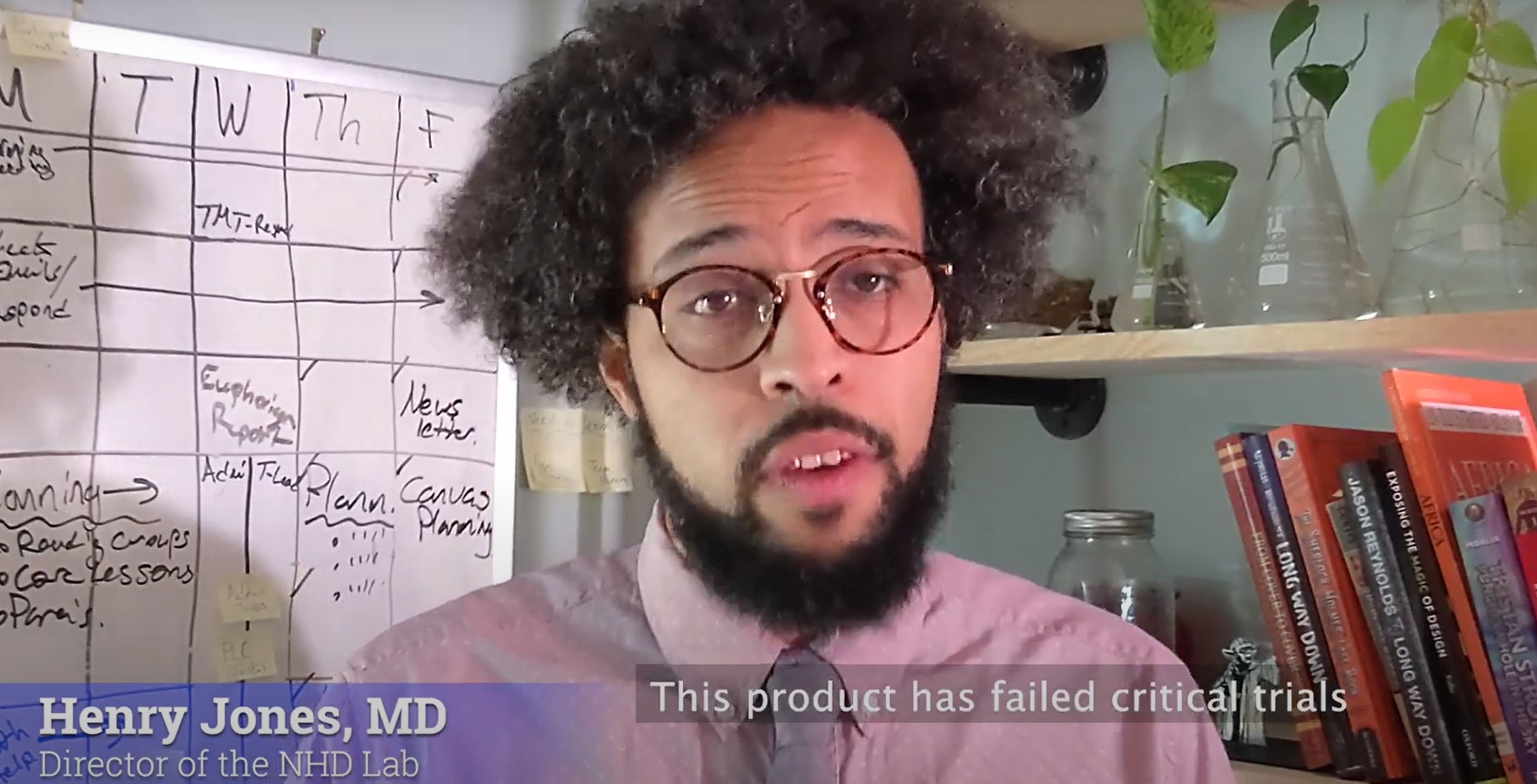}
  \Description{A screenshot of the deepfake video from Euphorigen, in which a doctor claims that the supplement failed critical trials and that the government will harm the population.}
  \caption{A screenshot of the deepfake video from Euphorigen, in which a doctor claims that the supplement failed critical trials and that the government will harm the population.}
  \label{deepfake}
\end{figure}

The narrative scenario of Euphorigen was designed by the Loki's Loop research team at the University of Washington, taking suggestions for narrative elements (e.g., big pharmaceutical company, playing as an investigator) from public librarians with experience developing and delivering escape rooms, and from students participating in a directed research group \cite{Turns_Ramey_2006} to support the project \cite{Cho_Coward_Lackner_Windleharth_Lee_2023}. The goal was to create a fictional narrative that differed enough from the political misinformation examples that were commonly being reported after the national election, so that the game would attract players regardless of their political inclination. An escape room developer was hired to create the puzzles, which were reused for Galaxy. In the narrative, players are initially tasked with investigating suspicious claims about the government putting a particular supplement into the national water supply. They are led into spreading a video claiming that the supplement is dangerous (Figure \ref{deepfake}). However, players then discover that this video is a deepfake, and have to find and share the original video which states that the supplement is safe and effective. 

\subsubsection{Galaxy} 
\begin{figure}
  \centering
  \includegraphics[width=\linewidth]{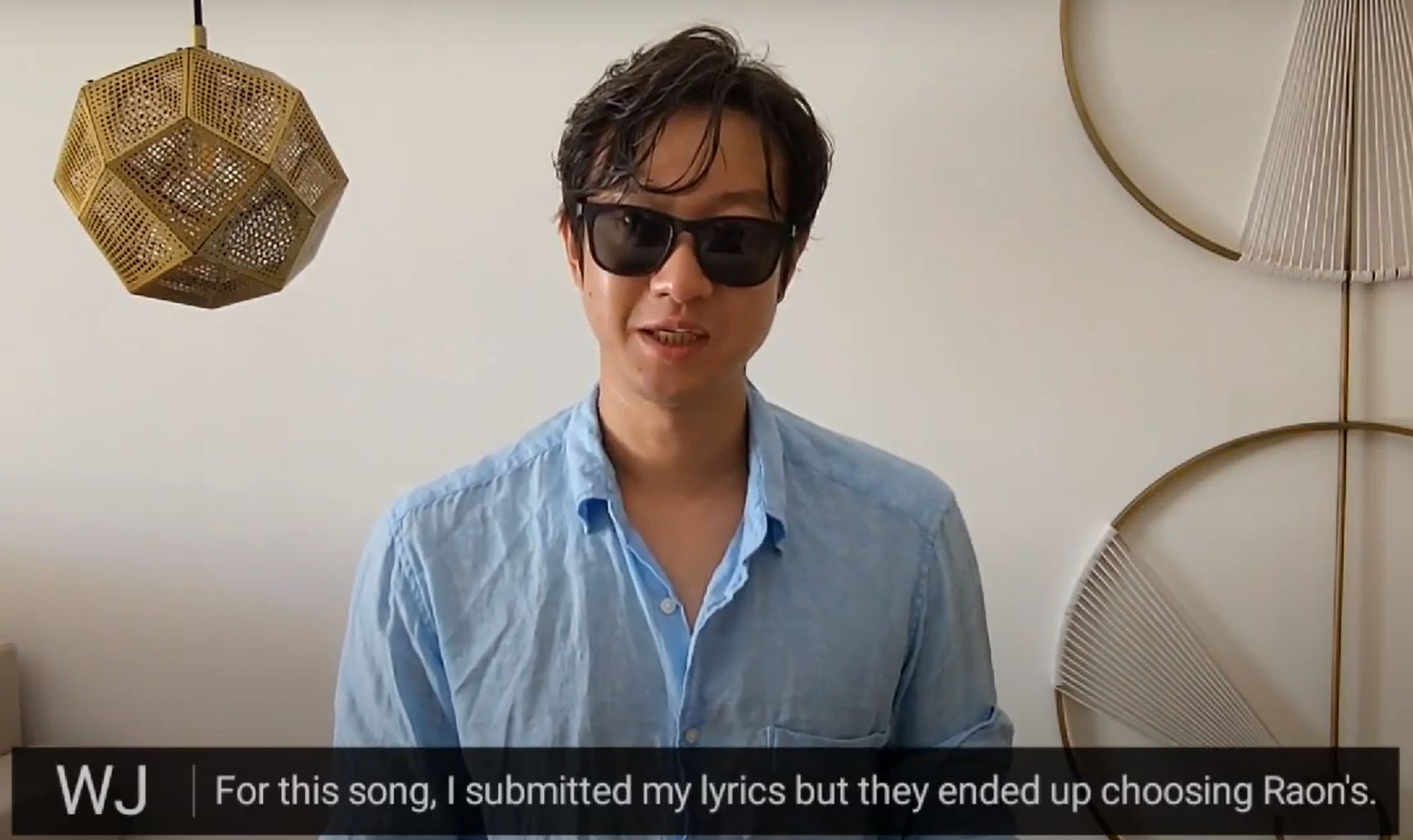}
  \Description{A screenshot of the deepfake video from Galaxy, in which WJ claims that the management company is giving other members of the band preferential treatment and that other members are not supportive of his solo work.}
  \caption{A screenshot of the deepfake video from Galaxy, in which WJ claims that the management company is giving other members of the band preferential treatment and that other members are not supportive of his solo work.}
  \label{johnny}
\end{figure}

Galaxy was co-designed with five participants who identified as members of the ARMY fandom, supporting the music group BTS. In previous work, we found that ARMY's tactics for dealing with the spread of rumors within their community can be a basis for effective grassroots efforts and strategies to build collective resilience to misinformation \cite{Lee_Santero_Bhattacharya_May_Spiro_2022}. The narrative scenario of Galaxy is based off certain rumors involving a Kpop music group. In the story, players are tasked by a fellow fan with investigating suspicious claims that their bias\footnote{A term normally used to refer to a favorite member of a group in popular culture} WJ, a member of a popular Kpop group, is being mistreated by the group's management company and other members. They are led into spreading a  video in which WJ claims that the management company and other members are in fact mistreating him, and he is eager to go on a solo tour (Figure \ref{johnny}). However, players then discover this video is a deepfake, and must find and share the original video, in which WJ speaks of being treated well by the management company and his excitement to go on tour with the group. 

\subsection{Study Conditions} \label{conditions}
\begin{figure}
  \centering
  \includegraphics[width=\linewidth]{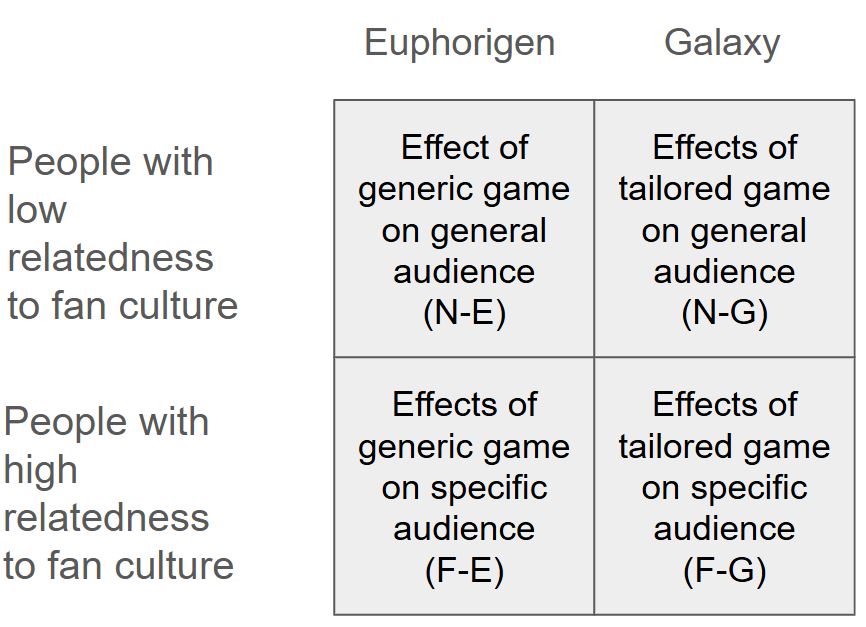}
  \Description{Diagram outlining the four study groups.}
  \caption{Diagram outlining the four study groups. The group shorthands are provided in parentheses, and will be used to refer to the groups in the following sections.}
  \label{studydesign}
\end{figure}

Based on their responses to the two screening survey questions involving fan engagement ("How often do you engage with online fan culture, particularly with specific figures/groups? (e.g., BTS, Taylor Swift) Engagement includes: following their social media accounts, engage with other fans, etc." on a 1-5 Likert scale, and "If yes, what popular figures/groups do you engage with?"), the primary researcher divided participants into two groups: non-fans and fans. Non-fans reported their fan culture engagement to be either 1 or 2 on a Likert scale, and did not specify any fan groups that they followed. Fans reported their fan culture engagement to be 3-5 on a Likert scale, and specified figures or groups that they engage with. Each group was further split into halves, where one half played Euphorigen and the other played Galaxy (see Figure \ref{studydesign}). Descriptive statistics for each study condition are summarized in Table \ref{tab:demographics}.

\begin{table}
  \caption{Descriptive statistics of each of the four study groups.}
  \label{tab:demographics}
  \centering
  \begin{tabular}{|c|c|c|c|c|}
    \hline
    Group & N-E & N-G & F-E & F-G \\
    \hline
    $n$ & 31 & 31 & 29 & 32 \\
    \hline
    Age & 24.16 & 22.16 & 22.83 & 22.77 \\
    \hline
    Gender (M/F/NB) & 13/17/1 & 8/22/1 & 6/22/2 & 5/23/3 \\
    \hline
    Fan Culture Engagement & 1.48 & 1.74 & 3.7 & 3.45 \\
    \hline
    Political Affiliation & 2.58 & 2.42 & 2.17 & 2.41 \\
    \hline
    Visits Social Media & 5.35 & 5.77 & 5.93 & 5.67 \\
    \hline
    Shares on Social Media & 2.97 & 3.29 & 3.03 & 3.19 \\
    \hline
    Visits News Media & 3.65 & 4.1 & 4.46 & 3.77 \\
    \hline
    Shares News Media & 2.26 & 2.71 & 2.97 & 2.81 \\
    \hline
  \end{tabular}
\end{table}

\begin{figure*}
  \centering
  \includegraphics[width=\textwidth]{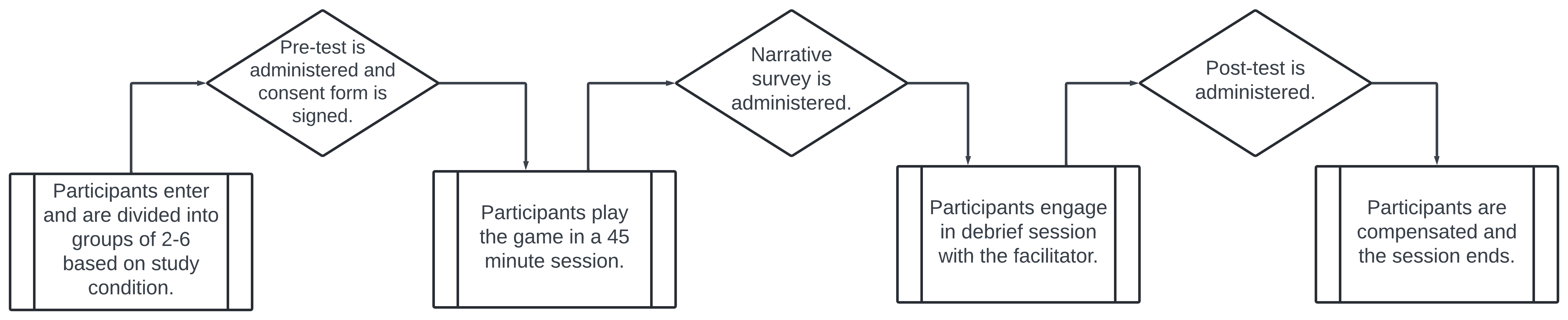}
  \Description{A visual representation of the session protocol; the diamonds represent the measures in sequential order of administration.}
  \caption{A visual representation of the session protocol; the diamonds represent the measures in sequential order of administration.}
  \label{session}
\end{figure*}

\subsection{Data Collection and Measures}
During the study session, participants were administered a pre-post session questionnaire as well as a post game narrative survey measuring transportation and identification. These are described below, and were administered in the sequential order presented in Figure \ref{session}.

\subsubsection{Pre/post questionnaire} Upon entering the study session, all participants were asked to fill out a questionnaire with four Likert scale questions regarding their perceptions of misinformation (ranging from 1 being `Not at all' to 7 being `Extremely'). We developed these questions during initial pilot testing of Euphorigen, and they are used to gain a baseline understanding for participants' opinions on misinformation and how this might affect their reactions to our games. Participants filled out the same questionnaire at the end of the session. The questions were:
\begin{itemize}
    \item I am concerned about misinformation's effect on society.	
    \item I feel confident in my ability to identify misinformation. 	
    \item I am concerned about accidentally believing misinformation. 
    \item I am concerned about accidentally sharing misinformation. 	
\end{itemize}

In the post-test, participants were additionally asked about their game experience using the same Likert scale in the following questions:
\begin{itemize}
    \item How much did you like the game?
    \item How much would you recommend the game to other players?
\end{itemize}

\subsubsection{Narrative measures}
Immediately after finishing the gameplay, participants were administered a survey assessing the following measures:
\begin{itemize}
    \item \textit{Transportation}: To measure narrative transportation, we used 10 out of 11 general items used by Green and Brock (2000) \cite{Green_Brock_2000}, excluding one item that reduces reliability \cite{Cohen_Tal-Or_Mazor-Tregerman_2015}. Each item was measured on a Likert scale from 1-7 (1 being `Strongly disagree' and 7 being `Strongly agree'), resulting in a total transportation score measured out of 70. We adapted the questions to refer to the game narrative rather than a generic narrative (e.g., "While I was playing the game, I could easily picture the events in the game's story taking place"). Transportation was measured in a questionnaire administered immediately after completion of the game. The reliability of the scale was excellent ($\alpha$ = 0.99). 
    \item \textit{Identification}: To measure narrative identification, we adapted six items from Cohen (2001): items 1, 3, 4, 5, 6, and 8 \cite{Cohen_2017}, excluding questions intended for evaluating intra-item validity. Each item was measured on a Likert scale from 1-7 (1 being `Strongly disagree' and 7 being `Strongly agree'), resulting in a total identification score measured out of 42. We adapted the questions to refer to the game narrative rather than a generic program (e.g., "I think I have a good understanding of the player character"). Identification was measured in a questionnaire administered immediately after completion of the game. The reliability of the scale was good ($\alpha$ = 0.897). 
\end{itemize}
\subsection{Session Protocol} 

Upon entering the session, all participants were asked to complete the consent form and pre-questionnaire. Then, participants were separated into groups ranging from 2-6 people, depending on session attendance. All members of a given group belonged to the same study condition. The session administrator provided groups with the game packets detailing the narrative scenario, and started a timer for 45 minutes. In the first phase of the game, players solved three puzzles in which they identified misleading charts, headlines, and AI-generated faces (Figure \ref{gameplay}). Providing these three answers to the administrator allowed players to proceed to the second phase, in which they solved a math puzzle. This answer unlocked a URL which they could access on their phones or laptops, leading them to a video stirring up controversy (e.g., Figure \ref{deepfake}). Sharing this video revealed to players that it was actually a deepfake with misinformation, and unlocked the final phase of the game, where players solved a graph puzzle displaying how the misinformation they shared had spread. The answer to this puzzle yielded a final URL, linking to the real video clarifying the misinformation, and sharing this video revealed the final passcode marking the successful completion of the game. After providing this to the administrator, players were asked to fill out the questionnaire with the transportation and identification measures. The administrator then ran the debrief session and asked players to complete the post-questionnaire before leaving the session. See Figure \ref{session} for the entire protocol flow.

\subsubsection{Debrief} After completion of the game and the questionnaire administering the transportation and identification measures, players participated in a debrief with the first or second author. The debrief was intended to facilitate a discussion with the players, explaining the game and its design intentions while eliciting player opinions and peer-to-peer sharing. The debrief questions were as follows:
\begin{enumerate}
    \item How was your experience? How did you feel? What did you learn?
    \item This game involved misleading headlines, manipulated charts, social media bots, deepfake images, and deepfake videos. Was there anything new or surprising to you about these tactics? 
    \item Have you ever believed something on social media that turned out to be untrue? Why do you think you fell for it and how did it make you feel?
    \item What are the consequences of sharing misinformation and how would it affect you or the people around you? Feel free to share a situation when you unintentionally shared information that turned out to be false.
    \item Were there any characters and story elements that you think are relatable?  What are they and why are they relatable? 
    \item In the game’s final puzzle, you were able to contact everyone who spread the deepfake video. How do you think this would have played out in real life? 
\end{enumerate}

\subsection{Hypotheses}
As Galaxy was co-designed with members of a fan community, participants who identify with fan communities may find it more relatable and tailored to their misinformation contexts than Euphorigen.
Conversely, non-fans playing Galaxy are unlikely to find the scenario very relatable as it is very specific to a fan culture context. They are more likely to find Euphorigen, which presents a more generic misinformation context, relatable. However, when compared to each other, both groups will likely have similar experiences playing Euphorigen, as all members of the study are part of the general audience for which the game was designed.
\begin{itemize}
        \item \textbf{$H1_T$}: Fans playing Galaxy will experience significantly higher transportation than fans playing Euphorigen. 
        \item \textbf{$H1_I$}: Fans playing Galaxy will experience significantly higher identification than fans playing Euphorigen.
        \item \textbf{$H2_T$}: Fans playing Euphorigen will experience similar transportation to non-fans playing Euphorigen. 
        \item \textbf{$H2_I$}: Fans playing Euphorigen will experience similar identification to non-fans playing Euphorigen.
        \item \textbf{$H3_T$}: Non-fans playing Euphorigen will experience significantly higher transportation than non-fans playing Galaxy.
        \item \textbf{$H3_I$}: Non-fans playing Euphorigen will experience significantly higher identification than non-fans playing Galaxy.
        \item \textbf{$H4_T$}: Non-fans playing Galaxy will experience significantly lower transportation than fans playing Galaxy.
        \item \textbf{$H4_I$}: Non-fans playing Galaxy will experience significantly lower identification than fans playing Galaxy.
    \end{itemize}

\subsection{Analysis} 

\subsubsection{Quantitative} Results from the screening, pre/post, and narrative measures surveys were filtered with Microsoft Excel and analyzed with Python. Python packages used were numpy, pandas, scikitlearn, and statsmodel.api. These enabled analyses such as Principal Component Analysis (PCA), and the creation of visualizations. ANOVA and t-tests were also conducted to answer our hypotheses. A Shapiro-Wilk test of the residuals showed that transportation was normally distributed (W(123) = 0.99, p = 0.564), but that identification was not normally distributed (W(123) = 0.98, p = 0.044). Levene's test further showed that the variances for transportation were not equal (F(3, 119) = 3.200, p = 0.026), and that variances for identification were equal (F(3, 119) = 2.173, p = 0.095). Hence, where appropriate, we used generalized linear mixed models (GLMMs) to estimate the linear relationship between the independent and dependent variables for both transportation and identification. We use ** to denote p < 0.05 and *** to denote p < 0.01. 

\subsubsection{Qualitative} The debriefs were audio recorded and transcribed using Rev.ai. Transcripts were then imported into Atlas.ti. The first and second author independently followed the six steps outlined in Braun \& Clarke \cite{Braun_Clarke_2006} to inductively code the transcripts and develop themes. They independently coded the first four transcripts, then combined these generated codes into themes. After a round of review, the authors split the transcripts in half and coded them using the developed themes. They then met once more to discuss the codes to consensus. 

\section{Results}
We answered RQ1 primarily through significance testing measures related to our hypotheses, and these results are presented in Section \ref{hypothesis}. RQ2 was answered through exploratory quantitative analysis; these findings are outlined in Section \ref{significance}. RQ3 was answered qualitatively, and the themes are outlined in Section \ref{qual}.

\subsection{Comparing Player Experiences Between Conditions} \label{hypothesis}

\subsubsection{Transportation and Identification}
Results for transportation and identification per study condition are reported in Tables \ref{tab:transportationmean} and \ref{tab:identificationmean}, respectively. As H1, H3, and H4 were directional, we used one-tailed two-sample t-tests to compare sample means between each study condition. We used a two-tailed two-sample t-test for H2, which was not directional. Results are summarized in Table \ref{tab:hypotheses}.

\begin{table}
  \caption{Mean and standard deviation for narrative transportation measure. Scores could range from 10-70.}
  \label{tab:transportationmean}
  \centering
  \begin{tabular}{|c|c|c|}
    \hline
    \textbf{Transportation}& Non-fan & Fan \\
    \hline
    Euphorigen & $41.58 \pm 7.69, n = 31$ & $39.41 \pm 8.03, n = 29$ \\
    \hline
    Galaxy & $33.42 \pm 10.63, n = 31$ & $40.16 \pm 6.81, n = 32$ \\
    \hline
  \end{tabular}
\end{table}

\begin{table}
  \caption{Mean and standard deviation for narrative identification measure. Scores could range from 6-42.}
  \label{tab:identificationmean}
  \centering
  \begin{tabular}{|c|c|c|}
    \hline
    \textbf{Identification}& Non-fan & Fan \\
    \hline
    Euphorigen & $26.74 \pm 7.09, n = 31$ & $28.55 \pm 8.63, n = 29$ \\
    \hline
    Galaxy & $24.22 \pm 7.69, n = 31$ & $30.97 \pm 5.70, n = 32$ \\
    \hline
  \end{tabular}
\end{table}

\begin{table}
\caption{Summary of t-testing our hypotheses.}
\centering
\begin{tabular}{|c|l|c|}
\hline
\textbf{Hypothesis ID} & \textbf{Null Hypothesis ($H_0$)} & \textbf{p-value} \\ \hline
$H1_T$ & $FG_T \leq FE_T$ & 0.34871 \\ \hline
$H1_I$ & $FG_I \leq FE_I$ & 0.09894 \\ \hline
$H2_T$ & $FE_T = NE_T$ & 0.24951 \\ \hline
$H2_I$ & $FE_I = NE_I$ & 0.37725 \\ \hline
$H3_T$ & $NE_T \leq NG_T$ & 0.00045*** \\ \hline
$H3_I$ & $NE_I \leq NG_I$ & 0.09272 \\ \hline
$H4_T$ & $NG_T \geq FG_T$ & 0.00226*** \\ \hline
$H4_I$ & $NG_I \geq FG_I$ & 0.0001*** \\ \hline
\end{tabular}
\label{tab:hypotheses}
\end{table}

Notably, two of the three significant results involve the low transportation and identification experienced by non-fans who played Galaxy. Indeed, these results also surfaced qualitatively in the debrief. Several participants from the Nonfan-Galaxy condition were skeptical that the scenario in Galaxy was realistic (e.g., "I don't think people would get this mad just because of a KPop band." - P29), though it was based off of a scenario that is common in the Kpop fandom. Other Nonfan-Galaxy participants were tangentially aware of such occurrences but had never experienced them personally (e.g., "The social media Kpop scene is very active on Twitter. So I've seen a lot of misinformation or people just wanting to make up stuff for clout. Even though I don't personally use Twitter, I already feel the impact and it seeps into YouTube and all these other social media." - P23). Many fans who played Galaxy, however, stated in the debrief that they found the narrative very relatable, and some participants even recognized different fandom events that the game story reminded them of. For example, P16 in the Fan-Galaxy condition stated: "The premise of all of the Kpop fans mobilizing on behalf of a mistreated band member is a direct echo to a Kpop fan history of 2011. And reading it, I felt like I had been that [person]. I remember talking to my friends like, can you believe how this band member is being treated? Meanwhile, I'm getting this information via some Korean fan through multiple layers of translation. And so who knows if I knew anything that was correct in that situation, but I definitely felt like [Galaxy's story] was extremely relatable." 

As expected, we failed to reject the null on H2, implying that both fans and non-fans related to Euphorigen similarly, being part of the same general audience. However, while we expected fans to relate significantly more to Galaxy than Euphorigen, the results were mixed. Though Euphorigen's misinformation context may not be tailored to people who frequently use social media like the fans do, it was likely still relatable enough to their personal experiences. 

\subsubsection{Game Effects on Pre-Post Tests} \label{prepost}
Our expectation was that scores for all the pre-test questions would significantly increase in the post-test; that is, confidence in identifying misinformation would increase, but concern about misinformation would also increase, based on the learning goals of the game and previous pilot testing. We performed one-tailed paired t-tests within each of the four study conditions, as well as by game and by non-fans/fans. To compare pre/post results between different groups, we used two-tailed, two-sample t-tests.

For Q1 (I am concerned about misinformation's effect on society), concern rose significantly among players of Galaxy (5.79 $\pm$ 1.23 to 6.02 $\pm$ 1.17; p = 0.019**). This may have been affected in part by the significant increase in concern specifically among the non-fan players of Galaxy (5.68 $\pm$ 1.42 to 5.97 $\pm$ 1.30; p = 0.036**). Exposing non-fan participants to a misinformation context that they normally do not interact with may have made them realize that misinformation they previously did not find particularly harmful can actually have significant impacts. For Q2 (I feel confident in my ability to identify misinformation), fans' confidence in identifying misinformation increased significantly (4.51 $\pm$ 1.08 to 4.78 $\pm$ 1.03; p = 0.049**). While participants from all conditions expressed appreciation for the puzzle that taught them about hallmarks of AI-generated humans, these learnings may have been more relevant to fans, who interact with falsified images of idols in their fandoms more than non-fans do. We saw no significant results across any study subgroups for Q3 (I am concerned about accidentally believing misinformation); participant concern was already fairly high in the pre-test, making large changes in the post-test unlikely. Finally, we similarly saw no significant results for Q4 (I am concerned about accidentally sharing misinformation).

\subsubsection{Enjoyment and Recommendability}

Our post-test additionally included two questions on player experience. These questions asked about player enjoyment (Table \ref{tab:pxenjoyment}) and whether or not they would recommend the game to others (Table \ref{tab:px_recommend}), on a Likert scale from 1 to 7 (ranging from 1 being `Not at all' to 7 being `Extremely'). To compare results between groups, we used one-tailed, two-sample t-tests. We found moderately positive correlations between transportation/identification and enjoyment ($r(113) = 0.5317, p < 0.001; r(113) = 0.4825, p < 0.001$) as well as transportation/identification and willingness to recommend the game ($r(113) = 0.4351, p < 0.001; r(113) = 0.4341, p < 0.001$). The literature also supports the assumption that higher levels of transportation and identification have been found to promote higher game enjoyment \cite{crutzen2016enjoyment, Hefner_Klimmt_Vorderer_2007}. Furthermore, there was strong correlation between levels of enjoyment and recommendability ($r(116) = 0.8277, p < 0.001$). As such, we had similar expectations surrounding player experience as we did regarding our hypotheses, and centered our analysis around the following inquiries: 
\begin{itemize}
    \item Did players in the fan condition enjoy/recommend Galaxy more than fan players of Euphorigen? (corresponding to H1)
    \item Did players in the non-fan condition enjoy/recommend Euphorigen similarly to fan players of Euphorigen? (corresponding to H2)
    \item Did players in the non-fan condition enjoy/recommend Euphorigen more than non-fan players of Galaxy? (corresponding to H3)
    \item Did players in the fan condition enjoy/recommend Galaxy more than non-fan players of Galaxy? (corresponding to H4)
\end{itemize}

In terms of enjoyment, players in the fan condition enjoyed Galaxy slightly more than Euphorigen, but it was not significant. However, players in the non-fan condition enjoyed Euphorigen significantly more than Galaxy (p = 0.012**). This aligned with what we observed in our hypothesis testing of player levels of transportation and identification in H3. However, despite the fact that players in the non-fan condition reported significantly lower levels of transportation and identification in Galaxy than fans did, they did not enjoy the game significantly less than fans. This could be explained by separating the narrative component, which concerns transportation and identification, from the rest of the game mechanics. Participants likely found the puzzles to be entertaining on their own, and enjoyed the experience of solving them with a group. 

Deviating from our original assumptions surrounding H2, players in the non-fan condition enjoyed Euphorigen significantly more than players in the fan condition did (p = 0.006***). This was surprising as we had not observed any significant differences in transportation or identification between fan vs. non-fan players of Euphorigen. A potential explanation is that while fans may have felt similar levels of transportation and identification to non-fans, their experiences may have differed affectively, especially since fans provided their many negative encounters with misinformation in the debrief. Some fans in the Euphorigen condition may have also had inherent preferences for more tailored contexts, despite not knowing about Galaxy. For example, P13 stated, "I think I believe dumb pop culture things more than...vaccine headlines. I remember a couple months ago there was...an alleged picture of Taylor Swift and Travis Kelce on a swing together or something. And I was like, oh my God, that's so cute. And then later I found out that it wasn't really them and I was like, whoops."

\begin{table}
  \caption{Mean and standard deviation for reported game enjoyment.}
  \label{tab:pxenjoyment}
  \centering
  \begin{tabular}{|c|c|c|}
    \hline
    \textbf{Game enjoyment}& Non-fan & Fan \\
    \hline
    Euphorigen & $6.35 \pm 0.55$ & $5.76 \pm 1.15$ \\
    \hline
    Galaxy & $5.78 \pm 1.25$ & $6.13 \pm 0.95$ \\
    \hline
  \end{tabular}
\end{table}

\begin{table}
  \caption{Mean and standard deviation for how inclined players felt to recommend the game to others.}
  \label{tab:px_recommend}
  \centering
  \begin{tabular}{|c|c|c|}
    \hline
    \textbf{Recommendability}& Non-fan & Fan \\
    \hline
    Euphorigen & $6.25 \pm 0.73$ & $5.79 \pm 1.05$ \\
    \hline
    Galaxy & $6.04 \pm 1.09$ & $6.31 \pm 0.89$ \\
    \hline
  \end{tabular}
\end{table}

In terms of recommendability, fans were significantly more likely to recommend Galaxy over Euphorigen (p = 0.023**), aligning with H1. However, non-fans were not significantly more likely to recommend Euphorigen over Galaxy, despite enjoying it significantly more. This likely also affected the finding that fans were not significantly more likely to recommend Galaxy than non-fans. As previously discussed, although players in the non-fan conditions are themselves not involved in online interest groups, they know people who are, or have seen others on the internet who would relate to fan contexts, which is a potential explanation for the Nonfan-Galaxy condition's relatively high inclination to recommend Galaxy to others. For example, as P19 from the Nonfan-Galaxy condition stated, "I'm not involved in any fandoms, but I see this kind of thing play out all the time as like a bystander...I think it played out about as how it does in real life too...people get their hands on misinformation and it's inflammatory so they spread it without verifying and then even if the truth comes out, it's often less distributed or less exciting so it doesn't reach people the same way."  

Furthermore, once again, our expectations surrounding H2 were not met: non-fans were significantly more likely to recommend Euphorigen than fans (p = 0.024**). As the correlation between levels of enjoyment and recommendability was strong, similar reasoning from the previous section may also apply to this result. 

\subsection{Factors Affecting Transportation and Identification} \label{significance}
As the variances for transportation were not equal, we used Welch's ANOVA, which revealed that playing Galaxy had significant effects on transportation (F(1, 120) = 5.930, p = 0.016**), likely due to the low mean transportation experienced by non-fans who played Galaxy. A two-way ANOVA for identification showed that fans reported higher identification than non-fans regardless of game played (F(1, 120) = 10.572, p = 0.0014***). This may be due to fans' slightly higher involvement with online spaces and misinformation in general (see Table \ref{tab:demographics}). 

A key point in both games was the sharing of misinformation in the form of a deepfake video. We investigated if reported frequency of sharing on social or news media had an effect on transportation and identification. Overall, reported frequency of sharing news media had significant effects on both transportation (p = 0.005**, GLMM) and identification (p = 0.045**, GLMM). Other relevant factors to Galaxy's plot, such whether participants used Twitter or followed Kpop groups, had no significant effects on either transportation or identification. Similarly, among those who played Euphorigen (n = 60), in which a key plot point involves investigating news misinformation, reported frequency of sharing news media had a significant effect on identification (p = 0.026**, GLMM). Reported levels of visiting social or news media had no significant effects on transportation or identification in either game. This may be because a majority of young adults are aware of misinformation on social media independent of how often they visit social media themselves. 

Upon investigating if answers to the pre-test had significant association with either transportation or identification, we found that reported concern about sharing misinformation (Q4) had a very significant effect on transportation (p = 0.009***, GLMM) and was positively correlated, indicating that participants with pre-existing higher levels of concern found the game narratives to be more believable.

\subsection{Analyzing Player Debriefs} \label{qual}
The player debriefs helped us develop a qualitative understanding of player experiences in the game, as well as their broader interactions with misinformation narratives in their daily lives. These findings explore young adults' interactions with misinformation, revealing widespread encounters and high concern about its impact. Participants recalled instances of believing misinformation, from pop culture rumors to politically charged falsehoods, often amplified by AI-generated content. Emotional investment emerged as a key factor in susceptibility, particularly in fandom contexts where misinformation can intensify community dynamics. While most admitted to falling for misinformation, fewer believed they had shared it. However, fans were more open about sharing misinformation in group chats, driven by urgency or excitement. We highlight how emotional and contextual factors shape young people's engagement with misinformation, with trust applied selectively depending on the topic’s perceived consequences.

\subsubsection{Interactions with misinformation} Unsurprisingly, every participant in this study had interacted with misinformation before, and overall concern regarding misinformation in the pre-test was high ($5.79 \pm 1.24$, n = 123). When prompted to recall instances in which they had believed misinformation themselves, participants provided examples ranging from falsified pop culture rumors, e.g., that celebrities such as Jaden Smith and Betty White had died (P103, P15), to politically charged misinformation surrounding recent current events. For example, P17 discussed how "during COVID, there were a lot of social issues that people were talking about on Instagram...[and they] definitely ended up taking a side of some polarizing issue that [they] didn't know enough information about to be taking the right side." P69 also mentioned that "the misinformation with the Israel-Palestine conflict [had] ten different versions from both sides. It's happening so fast and so frequently with multiple stories from all perspectives that...even after the fact, [they didn't] know really which one was true in reality." Many participants also explicitly mentioned Photoshopped or AI-generated media that they had fallen for, e.g., Photoshopped album covers misleading viewers into believing there was a new release (P15), AI-generated song covers (P107), or Photoshopping two celebrities together to fuel dating rumors and gain clout from virality (P16). Several participants could not recall particular examples of falling for misinformation when prompted, but acknowledged that it was very likely that they had. As P17 put it, "I feel like inevitably, at some point, I have."

\subsubsection{Consequences of misinformation}

On an individual level, several participants expressed the embarrassment of falling for or sharing misinformation, or being personally upset that they accidentally contributed to something potentially harmful. For example, as P16 stated: "It's also embarrassing when we reach that point in the game and that we had no option other than to share, but there's the little feeling of, hmm, maybe this is gonna turn out not to be true. It feels bad to be spreading this information, but also it feels embarrassing if you have fallen prey to it too". They also acknowledged that "it also tarnishes your reputation, if you share misinformation and then the other person recognizes that it's misinformation" (P22). More broadly, participants reflected on how misinformation "definitely affects relationships and people, and the way people perceive each other" (P16), "creates vaccine hesitancy" (P18), "confirms different stereotypes" (P22), and how it "just becomes a cycle and then nobody knows what the real information is" (P77). Indeed, it was notable that several of the young adults in the study expressed "deep skepticism of everything [they] see on the internet" (P23), with many stating that they had mostly quit social media altogether. As P29 strongly stated, "What's the point of social media? It's supposed to make information more accessible, but it's actually just easier to spread. I've gotten to the point where I don't even care about social media and news outlets, or look at people and their opinions on social media because so many things are fake on there." P30 chimed in, stating, "It's really scary because I would say a good chunk of the internet right now is misinformation and a lot of people who have limited digital exposure or literacy don't even have pathways to identify it." 

\subsubsection{Why young people are falling for misinformation} \label{falling}

In spite of the paranoia that many participants brought to their social media experiences, nearly all of them discussed instances in which they had fallen for misinformation, and why this might be. The first reason was a lack of media literacy around AI-generated images, audio, and videos. Participants expressed that the rise of AI technologies has made it outright difficult to identify what may and may not be misinformation, likely contributing to their overall skepticism for social media. Almost every player group discussed how difficult identifying the AI generated images was in the game puzzle, and many did not know the strategies the game provided as typical hallmarks of a generated human (e.g., mismatched or asymmetrical accessories, extra teeth, distortions in background or hair). P4 stated that the game made them realize that "[they] don't really fact check pictures and videos as much as [they] do articles, [and] tend to assume that if it's a picture it must be real. If it's realistic, [they] don't generally take the time to check it". P19 referred to generative AI technologies as "a force multiplier that makes things more accessible too...for Photoshop, you actually needed skill for people to believe it". P25 referenced their answer to the pre-test, stating, "When I said my confidence was 5 out of 7 for recognizing misinformation, I wasn't including deepfakes. I don't think I can recognize those." 

However, the majority of the reasons that participants believed they fell for misinformation were more personal. Interest and investment were cited as factors for being more prone to believing misinformation, and this was especially discussed by members of the fan condition. Many of these players were Kpop fans themselves, and they provided several examples of instances where they thought their fandom's investment into an idol or figure went too far, e.g., line distribution scandals trending on Instagram (P64) or trucks sent to protest at management companies over idol dating rumors (P112, P115). As P56 put it, "The more that you have an emotional response to something, the harder it is to really see it for what it could be. And if it's misinformation that you're just very excited or really afraid of, you're more likely to trust it." P110 further pointed out that "especially in these situations, it's hard to counter something with logic because people are very emotionally invested". 

In addition, players provided examples both from non-fan and fan contexts of how emotional investment in a topic can be preyed upon by bad actors in the misinformation sphere. P111 shared that "certain actors or idols will make up rumors and gossip to ruin the careers of idols they don't like, and get their fans to spread it". Outside of fan contexts, many participants shared examples from current political events. P119 discussed the fear and uncertainty around certain politically charged misinformation: "I'm from Korea and if there's [sic] fake videos made with the president of North Korea and [it says] we might be attacking, or something about missiles, even if it's fake, we'll freak out." P100 expressed regret about family dynamics that have changed in light of misinformation surrounding the Israel-Palestine conflict: "The misinformation there is actually horrifying. That's hard, because I've known a lot of these [family members] my whole life and they've always been so kind, so caring...but then this happened and now I can't open Facebook. You can't even talk to them about it either because they're so set in it. It's like they feel obligated to have the opinion and it's scary."

Finally, despite the distrust that many of our participants expressed towards social media ecosystems, participants discussed how they selectively applied skepticism based on how harmful they perceived the information to be. We discuss this in depth in the next section. 

\subsubsection{Perceived importance of information affects believing and sharing behaviors}

Although nearly all participants stated that they had fallen for misinformation, fewer believed that they had shared misinformation themselves. This was reflected in the pre-test answers as well: participants were less concerned about sharing misinformation ($4.71 \pm 1.68$, n = 123) than they were about believing misinformation ($5.05 \pm 1.34$, n = 123). As sharing misinformation was a key plot point in both Euphorigen and Galaxy, participants brought this up often as a point of narrative dissonance, stating that they “wouldn't have shared [the video] if it hadn't been necessary to our progress in the game” (P23). As P24 explained, “I don’t share…and definitely not in this context where we were already aware of the fact that it is potentially fake" (P24). Some non-fan participants who played Galaxy were skeptical  that fans would be so willing to share anything about their fandom, e.g., P29, who stated, "I would say that the actual scenario here seems really unrealistic for me. That you're a super fan of something and you just have to share any information you find about it." However, many fan players openly admitted to public sharing of misinformation related to their fandom, the best example of which was given by P70: "If I see something, it's going to all my group chats right away, to multiple different people. Even if I see something that might be a deepfake, I'm still going to share it to people thinking, oh my God, something is happening, you should go check it out before it gets taken down. Especially some of the Taylor Swift things, some people didn't even get to see it before it got taken down." 

This quote is perhaps representative of an overarching phenomenon we observed through our discussions with participants: that identifying the difference between legitimate information and misinformation is only necessary if the topic is deemed consequential. As P2 explained, "I think it depends on the severity and consequences of the information…If it's a life changing piece of news, regardless of whether it's actually true or not, I'm going to be more distrustful of it. The consequences of believing something like that versus some fun fact, or celebrity gossip, are [much worse]". An additional factor in participants’ internal judgments was where they might be sharing misinformation. While fan groups were the most public venues that participants discussed sharing misinformation themselves, several participants discussed that although they might not share on social media, they were more likely to share to friends, family group chats, or through in-person conversations.

\section{Discussion}

Our findings showed that our initial hypotheses about transportation and identification in the two misinformation narrative games were either met (H2, H4) or partially met (H1, H3). Other factors, such as reported engagement with fan culture, social/news media sharing behaviors, and pre-existing concern for misinformation, also significantly affected transportation and identification. This provides sufficient evidence that the narrative was not only perceived differently between different player groups, but also had an effect on their gameplay experiences in a misinformation education context. Qualitatively, we found that the young adults in this study were well aware of standard media literacy skills and strategies for dealing with misinformation, such as the importance of choosing reliable sources and verifying information that they encounter. They were also aware of their shortcomings regarding AI-generated media, and were generally suspicious of information on social media. However, as discussed in the debriefs about the games and their narratives, these traits do not prevent them from sharing misinformation that they might perceive as less important or not immediately affecting them, or sharing within personal contexts. In the following sections, we discuss the implications of this for future misinformation education interventions. 

\subsection{The story matters} The different misinformation contexts that participants brought to the study affected how they related to the game. Quantitatively, it was apparent that the non-fan condition was unable to connect with Galaxy's narrative in the same way as fans were, as shown by the significantly lower levels of transportation and identification, implying reduced engagement and lowered emotional impact. Fans identified with and enjoyed Galaxy slightly more than Euphorigen, showing that co-designing led to slightly improved engagement over the general game for this particular group, although it was not statistically significant. Qualitatively, however, was where the difference in narrative framing appeared to matter most. In the debrief, where we intended most of the discussion and learning to occur, fans readily connected Galaxy's story to scenarios in their own lives. This aligns with previous research which shows that player understanding is developed through cycles of performance within game worlds \cite{squire2006content}. While most participants were able to relate either story to general instances of misinformation they had encountered, this showed us that fans were not only able to quickly connect the scenario to their own lives, but also to a subject of personal interest. Notably, several non-fan participants who played Galaxy expressed confusion or skepticism at the scenario, or expressed that they did not personally relate, even if they had friends who might. 
 
Indeed, the importance of personal interest cannot be understated. As misinformation is a controversial topic, and the young adults in our study seemed to be extremely burnt out and skeptical of misinformation and social media, engaging them through personal interests is potentially more important than ever before. In a phenomenon deeply mediated by emotions and personal belief, a game that is well-designed and informative may not always be sufficient to sustain engagement if it does not resonate with the audience’s specific interests or lived experiences \cite{young2021role}.

Regardless of condition, all participants were able to experience a small dose of the emotional impact of sharing misinformation, as evidenced by the surprise and embarrassment many of them expressed upon learning that the video shared in the game was a deepfake. Our results showed that participants' reported frequency of sharing on social media had varying levels of significance on transportation and identification. From this, we can infer that the emotional impact of sharing a fake video, even in the game, was enough to have an effect on player experience. 

Sociocultural narratives are lacking in the current body of misinformation games, and many of them address the rational rather than emotional components of misinformation adoption \cite{Devasia_Lee_2024}. Indeed, this work speaks to the importance of something that we believe many current misinformation education games often miss: that emotions and personal investment in a misinformation context are highly relevant to what we choose and decline to believe. Despite participants' self-professed confidence in their own media literacy skills and ability to identify misinformation, several still discussed falling for it themselves. Understanding how to identify misinformation is not necessarily enough to prevent a person from falling for it, especially if the topic is of emotional or personal interest to them. Narrative serves as a useful mechanism to provide that context for players to experience misinformation in a more realistic scenario than a program primarily focused on teaching them media literacy skills. 

\subsubsection{Strengths and drawbacks of designing generalized misinformation education interventions} Euphorigen proved to be the more successful game in terms of transportation and enjoyment, demonstrating that any player familiar with American news contexts found it to be moderately believeable and entertaining. This speaks to the potential efficacy of well-designed generalized interventions, and for educators and designers with limited time and resources, a general intervention may serve well for running in classrooms, libraries, or workplaces. 

However, generalizing has its limitations. While all our participants were from a relatively homogeneous audience - primarily American students at our university - the game narrative may be difficult to generalize across cultures and languages. For example, Euphorigen's narrative, which plays upon the inherent distrust that Americans have for the government and pharmaceutical companies in order to discuss confirmation bias, is likely not applicable to countries with significantly higher indices of trust in their governments \cite{Finland}. In addition, while generally relatable to students of a large, liberal American university, Euphorigen's similarity to the COVID-19 vaccine rollout may alienate certain audiences prone to vaccine skepticism. Indeed, these games were designed to be implemented in public libraries, and librarians face difficulties in deploying such interventions to audiences who need them most due to the fact that misinformation issues are often highly polarized \cite{young2021role}. In the current day, one may argue that there is almost no singular piece of information that is universally agreed upon, making a truly general intervention challenging. Writing general scenarios that can be lightly tailored to fit the widest range of contexts is likely the most optimal method of reaching the largest audience. 

\subsubsection{Strengths and drawbacks of co-designing misinformation education interventions with specific user groups}

As shown by the numerous quotes of fans discussing how the topics from Galaxy's narrative were relevant to their own lives, a primary strength of co-designed narratives such as Galaxy's is that it allows players to interact with misinformation education through the lens of their personal interests. Not only does this increase the potential for higher engagement in the intervention itself, but as personal investment and emotion are fundamental players in misinformation adoption, targeted narratives are an opportunity to engage with people in contexts in which they may be more prone to misinformation. Our results showed that fans identified with, enjoyed, and would recommend Galaxy significantly more than Euphorigen, which is a strong case for co-design as a way to increase the appeal of misinformation education for specific target groups. 

A potential drawback of co-designed interventions such as Galaxy is over-optimizing to a specific community, as shown by the relatively low transportation and identification experienced by non-fans who played the game. Any intervention designed with a particular group of people in mind may run the risk of alienating players who do not identify with that group, as shown by our results. Furthermore, co-design can be time consuming and resource intensive for improvements that may not be highly significant over a generalized version; for researchers and designers optimizing for the largest reach, a generalized intervention such as Euphorigen may be preferable. 

\subsection{The dangers of not caring enough}
Barzilai \& Chinn \cite{barzilai2020review} discussed how in many cases, people are not committed to validating accuracy or impartiality in favor of other personal goals, such as promoting personal interests or aligning with family and friends. In our debrief discussions, we observed that young people are very skeptical of and careful about validating sources around information that they perceive to be of high personal importance. Examples primarily included political and medical misinformation; many participants stated that they would not have shared the deepfake video in Euphorigen without verifying it first, as it contained information of high potential impact to others. However, this sense of care did not apply to misinformation that they perceived as less important, as exemplified by several of the participant quotes discussed in Section \ref{falling}.

This points to a potentially troubling revelation for misinformation education: that the verity of information, regardless of the topic, only matters insofar as its perceived importance to a user. For example, though many non-fans who played Galaxy expressed incredulity at the narrative scenario presented, and several participants wondered why people would care about misinformation about a celebrity, the scenario was based off of real life events that caused harm to a person. This finding represents a potential future direction for misinformation education, which can use Barzilai \& Chinn's framing to interrogate why people may not care enough about the verity of certain types of misinformation. They suggest emphasizing the development of dispositional intellectual virtues, i.e., habits of mind that dispose people to good thinking. They also suggest fostering students’ intellectual identities and agency, which would aid students in bridging the gap between knowing the correct course of action and executing it \cite{lapsley2020post}. We believe that the narrative framing of misinformation interventions is crucial to encouraging the reflection needed to build such intellectual virtues, more so than specific game mechanics or modalities. 

\subsection{Further design considerations for misinformation narrative games} 

\begin{figure}
  \centering
  \includegraphics[width=\linewidth]{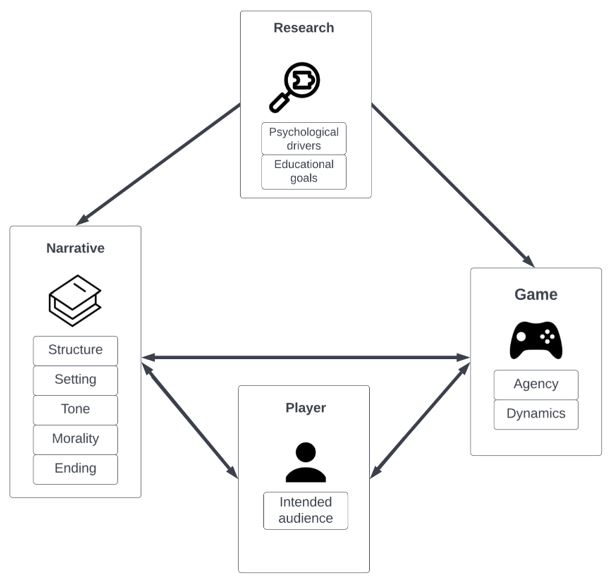}
  \Description{A summary of the Misinformation Game Narrative Design framework, presented in our previous work.}
  \caption{A summary of the Misinformation Game Narrative Design framework, presented in our previous work \cite{Devasia_Lee_2024}.}
  \label{mgnd}
\end{figure}

In previous work, we presented a framework outlining elements to consider when designing a narrative-centered misinformation intervention (see Figure \ref{mgnd}) which can serve as an additional theoretical lens to explore the role of narrative \cite{Devasia_Lee_2024}. This study highlights additional elements, beyond those discussed in the framework, that should be considered when designing narratives for misinformation education games. While this study focused on investigating the effects of a particular type of social media usage (i.e., engagement with fan culture) on perceptions of misinformation narratives, we may conjecture that broader cultural contexts are also likely to strongly affect the experience of playing such games (e.g., ethnicity, social norms), as evidenced by participants who originated from or have family internationally. For example, a narrative such as Euphorigen, which draws elements from vaccine skepticism, may not resonate with an Asian audience, where vaccines are more accepted \cite{sallam2021covid}. The recency of the misinformation context is also relevant to player experience; e.g., games about election misinformation would likely be most pertinent during an election cycle. Prior research also shows that different countries have different level of resilience to disinformation considering various indexes of polarization, populism, media trust, and media literacy \cite{humprecht2020resilience, Finland}, which can also impact narrative design. Another consideration is the physical context of the players; for example, if they are playing in an individual or group setting, and the social dynamics of who they may be playing the game with. Playing in certain settings, such as classrooms, workplaces, or libraries, may be time constrained to a class period or an activity block. Playing in groups with peers, as compared to playing with managers or subordinates in one's workplace, will likely lead players to make different choices. While we were not able to disambiguate the effects of social play within this study, it likely has an effect on how players perceive the narrative, although this has not been explored adequately in the literature. Regardless of a specific narrative or misinformation context, certain tactics are common across multiple contexts. For example, in both games, players praised the puzzle which taught them how to identify AI-generated images of people, as this tactic has become ubiquitous across social media and news platforms generally. Lastly, designing different kinds of post-game experiences, such as a resource kit, a quiz, or a debrief session can help reinforce the learning experience from the game.

\section{Limitations} This study was performed at a large American university which limits the level of diversity that could be achieved in terms of ethnicities or political affiliations. As the effects of misinformation vary strongly by individual demographics, a more holistic investigation of how these games and their narratives affect transportation and identification among different populations is needed. For example, while a version of Euphorigen was co-designed and piloted with BIPOC Americans, it was not investigated in this study. Another limitation was that despite the fact that Galaxy was co-designed with avid Kpop fans, we were unable to recruit for just Kpop fans specifically in this study, and our fan condition represents a somewhat broad coalition of fans that may not have found the narrative as relatable to them.

\section{Conclusion and Future Work} 
In this work, we investigated how narrative transportation and identification were experienced by two different player groups in two misinformation education games. We found significant differences between player groups, and additionally observed that online sharing behaviors and prior concern regarding misinformation affected narrative measures. We discuss drawbacks in current misinformation interventions as well as suggest future directions for cultivating stronger intellectual dispositions in young people who frequently interact with social media. We contribute an empirical understanding of how narrative framing impacts player experience in misinformation education games. 

We are working to adapt these games to different international contexts, localizing their contents such that they are culturally relevant to specific populations. We wish to investigate elements of narrative transportation and identification for the localized games, and compare to the original versions targeted at American audiences. For example, preliminary observations revealed that players in Scandinavia did not share the inherent mistrust American players have towards the government in Euphorigen, and therefore had different reactions towards the story. A deeper investigation into cultural and historical contexts where the game will be played may help us identify strategies for creating effective narratives for misinformation games. We are also currently developing a misinformation game targeted at young adults which is meant to highlight the potential dangers of sharing AI-generated misinformation that may seem unimportant and humorous at first glance, but leads to harmful outcomes in the game story. We anticipate that this game will help us address the dangers of "not caring enough", as discussed in \cite{barzilai2020review}, in the context of misinformation more effectively. 

%
\begin{acks}
This work was supported by the University of Washington's Center for an Informed Public and the John S. and James L. Knight Foundation, through award number G-2019-58788.
\end{acks}

\bibliographystyle{ACM-Reference-Format}
\bibliography{main}

\end{document}